# Theoretical study of the excitation function in the CN + $C_2H_6$ hydrogen transfer reaction. Effect of vibrational excitation.

Joaquin Espinosa-Garcia, Cipriano Rangel, and Jose C. Corchado*

Área de Quimica Fisica and Instituto de Computación Científica Avanzada de Extremadura, Universidad de Extremadura, Badajoz (Spain),

e-mail: corchado@unex.es

In our previous paper[1] entitled "Role of the vibrational and translational energies in the CN($v$) + $C_2H_6$($v_1$, $v_2$, $v_5$ and $v_9$) reactions. A theoretical QCT study" we have theoretically analysed the impact of reactant vibrational excitations on reactivity, with particular emphasis on mode selectivity, at two total energies of 9.6 and 20.0 kcal mol$^{-1}$. Note that this total energy is the sum of the initial collision energy plus the vibrational excitation, and in the vibrational ground-state the total energy and the initial collision energy obviously coincide. The following vibrational modes were analysed: $v_1$, $v_2$, $v_5$ and $v_9$. The modes $v_1$ (3011 cm$^{-1}$ ~8.6 kcal mol$^{-1}$) and $v_5$ (2996 cm$^{-1}$ ~ 8.6 kcal mol$^{-1}$) correspond, to the symmetric and asymmetric C-H stretching modes, respectively, and were chosen to study mode selectivity because they differ by only 15 cm$^{-1}$. The modes $v_2$ (1428 cm$^{-1}$ ~ 4.1 kcal mol$^{-1}$) and $v_9$ (1006 cm$^{-1}$ ~ 2.9 kcal mol$^{-1}$) correspond to ethane bending modes and were chosen to study the effect of bending excitations. In that paper, we showed that at these two total energies, providing a certain amount of energy as translational energy resulted in a slightly lower reactivity than providing the equivalent amount of energy as vibrational energy. This effect was more pronounced at low energies, which is a counterintuitive scenario in an 'early' transition state reaction.

During the revision process of the paper, a reviewer suggested that it would be very useful to show excitation functions (integrated reaction cross section vs. collision energy) in the case of both ground-state and vibrationally excited reactants to observe the effect of initial translational energy. In the response letter we agreed with the referee's suggestion, but noted that it was challenging to perform these calculations in the short time of a review process, as they represent a huge computational effort, and could be considered in future research in our group. These new calculations are presented and analysed in the present Corrigendum.

Bimolecular reactions represent fundamental processes that govern diverse phenomena observed across chemistry, physics and biology. A crucial aspect of the study and characterisation of these reactions is the concept of excitation functions, which represent the dependence of reaction cross section on the energy of the colliding reactants. The insights offered by this concept include an understanding of the threshold energies required for the initiation of reactions, the presence of energy barriers along the reaction coordinate and the efficiency of energy transfer mechanisms occurring during collision events. Excitation functions are often characterised by intricate features that reflect the interplay between potential energy surfaces, molecular geometries, and vibrational modes of the reactant species. At low collision energies, these functions may change gradually indicative of barrierless reactions where no energy barrier impedes the formation of products. Conversely, reactions with energy barriers manifest as excitation functions with distinct activation thresholds, reflecting the energy required to surmount the barrier and initiate the reaction. On occasion, excitation functions may exhibit oscillatory structures corresponding to the vibrational excitation of reactant molecules, energy-dependent reaction paths, or quantum-mechanical resonances. Furthermore, excitation functions serve as valuable probes for investigating reaction dynamics and mechanisms, shedding light on factors influencing reaction selectivity, branching ratios, and stereochemistry.

In order to gain insight into the dynamics of the $CN + C_2H_6$ gas-phase reaction, a full-dimensional analytical potential energy surface, named PES-2023,[2] was developed as a combination of valence bond and molecular mechanic terms and fitted to high-level ab initio calculations at the explicitly correlated CCSD(T)-F12/aug-cc-pVTZ level. This reaction presents very high exothermicity, -22.20 kcal mol$^{-1}$, and it is practically barrierless, with a barrier height of 0.23 kcal mol$^{-1}$, being an "early" transition state reaction. Based on this surface, quasi-classical trajectory (QCT) calculations were performed. Figure 1 depicts excitation functions, i.e., reaction cross sections versus collision energy, in the range 0.2-20.0 kcal mol$^{-1}$ for the vibrational ground-state and the different vibrational excitations analysed in the present paper, $v_1$, $v_2$, $v_5$ and $v_9$. With regard to the vibrational ground-state, the excitation function presents a minimum at approximately 1 kcal mol$^{-1}$, with a slight increase at higher energies and a pronounced increase at lower energies. This behaviour is independent on the counting method employed, namely considering all reactive trajectories or excluding reactive trajectories with final vibrational energy of each product, HCN and $C_2H_5$, below their respective zero-point energy. The V-shape form of the excitation function is characteristic of non-threshold reactions. The pronounced increase observed at lower energies can be attributed to the substantial increase in the impact parameter within this energy regime, which increases from 4.0 Å at 20.0 kcal mol$^{-1}$ to 6.5 Å at 0.2 kcal mol$^{-1}$. This behavior was previously described kinetic study of the reaction.[3] In this work, a V-shaped temperature dependence of the thermal rate constants with temperature in the wide range 25-1000 K was reported the variation, which reproduced the experimental evidence. All vibrational excitations by one quantum, $v_1$, $v_2$, $v_5$ and $v_9$, present a similar V-shaped profile (Figure 1), with the stretching mode excitations exhibiting a more pronounced effect. These excitations enhance reactivity across the entire collision energy range, from 0.2 to 20.0 kcal mol$^{-1}$, by factors between 2.6 and 1.3. The stretching mode excitations, $v_1$ and $v_5$, produce the most pronounced effects while the bending excitations, $v_2$ and $v_9$, have a comparatively smaller impact. Furthermore, it should be noted that the vibrational excitations result in a shift of the minimum to larger collision energies, between 1 and 4 kcal mol$^{-1}$ for the stretching and bending modes, respectively.

To further investigate the different capacity of an equivalent amount of energy as translation or vibration to enhance reactivity, Figure 2 plots the reaction cross sections versus the total energy in the range 0.2-20.0 kcal mol$^{-1}$, for the ethane vibrational ground-state and the $v_1$, $v_2$, $v_5$ and $v_9$ vibrational excitations. It is shown that for a fixed total energy the vibrational energy is more effective than the initial translational energy, with the exception of the $v_2$ bending mode at high energies, where both curves are similar. This result represents, a priori, a counterintuitive example, considering that this reaction is an example of an "early" transition state. It should be noted that a previous research from our group[4] indicated that a similar phenomenon was observed in $O(^3P) + CH_4$ reaction, which also presents an "early" transition state. In that work we showed that vibrational energy is less efficient than an equivalent amount of translational energy. The differing behaviours of these two "early" reactions, classified as such based on the saddle point location, are related to the barrier height, 0.23 vs 6.4 kcal mol$^{-1}$, respectively, as well as a series of factors analysed in our previous paper,[1] including strong coupling between different vibrational modes and significant vibrational energy redistribution within the reactants before collision. These factors give rise to an unphysical energy flow, which results in the loss of vibrational memory before the reactants collide.

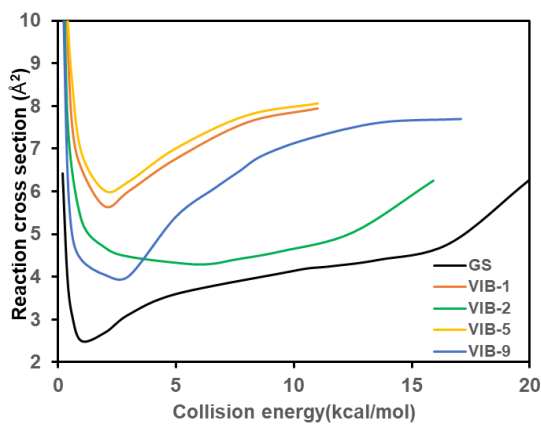

**Figure 1**. Reaction cross sections as a function of the collision energy in the range 0.2-20.0 kcal mol$^{-1}$ for the CN + C$_2$H$_6$($v_1$, $v_2$, $v_5$ and $v_9$) reactions. GS represents the ethane vibrational ground state

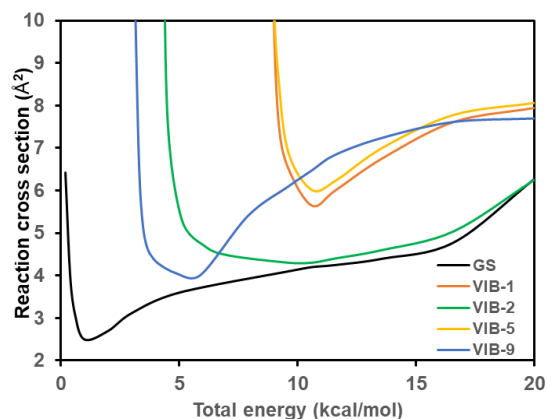

**Figure 2**. Reaction cross sections as a function of the total energy in the range 0.2-20.0 kcal mol$^{-1}$ for the CN + C$_2$H$_6$($v_1$, $v_2$, $v_5$ and $v_9$) reactions. GS represents the ethane vibrational ground state.